\documentclass[aps,prd,twocolumn,preprintnumbers,superscriptaddress,nobibnotes,floatfix,longbibliography,nofootinbib]{revtex4-1}

\pdfoutput=1
\usepackage{amsmath,amsfonts,amssymb,mathrsfs,graphicx,color,longtable}
\usepackage{hyperref}
\usepackage{graphicx}
\usepackage{amsfonts,amsmath,amssymb,bm}
\usepackage{array}
\usepackage{color}
\usepackage{enumitem}
\usepackage[explicit]{titlesec}
\usepackage[utf8]{inputenc}
\usepackage[normalem]{ulem}

\hypersetup{
     colorlinks   = true,
     citecolor    = red,
     urlcolor     = blue,
     linkcolor    = blue
}

\usepackage{xcolor}

\newcommand{\dd}{\mathrm{d}}

\begin{document}
\title{Type Ia supernovae data with scalar-tensor gravity}

\author{Mario Ballardini}\email{mario.ballardini@inaf.it}
\affiliation{Dipartimento di Fisica e Astronomia, Alma Mater Studiorum Universit\`a di Bologna, via Gobetti 93/2, I-40129 Bologna, Italy}
\affiliation{INAF/OAS Bologna, via Piero Gobetti 93/3, I-40129 Bologna, Italy}
\affiliation{INFN, Sezione di Bologna, viale C. Berti Pichat 6/2, I-40127 Bologna, Italy}
\affiliation{Department of Physics and Astronomy, University of the Western Cape, Cape Town 7535, South Africa}

\author{Fabio Finelli}
\affiliation{INAF/OAS Bologna, via Piero Gobetti 93/3, I-40129 Bologna, Italy}
\affiliation{INFN, Sezione di Bologna, viale C. Berti Pichat 6/2, I-40127 Bologna, Italy}

\begin{abstract}
We study the use of type Ia supernovae (SNe Ia) in the context of scalar-tensor theories of gravity, taking as 
a working example induced gravity, equivalent to Jordan-Brans-Dicke theory. Winking at accurate and precision 
cosmology, we test the correction introduced by a time variation of the Newton's constant, predicted by 
scalar-tensor theories, on the SNe distance modulus relation. We find that for induced gravity the coupling 
parameter is constrained from $\xi < 0.0095$ (95\% CL) using Pantheon SNe data alone down to $\xi < 0.00063$ 
(95\% CL) in combination with {\em Planck} data release DR3 and a compilation of baryon acoustic oscillations 
(BAO) measurements from BOSS DR12.
In this minimal case the improvements in terms of constraints on the cosmological parameters coming from the 
addition of SNe data to cosmic microwave background (CMB) and BAO measurements is limited, $\sim7\%$ on the 
95\% CL upper bound on $\xi$. 
Allowing for the value of the gravitational constant today to depart from the Newton constant, we find that 
the addition of SNe further tightens the constraints obtained by CMB and BAO data on the standard cosmological 
parameters and by 22\% on the coupling parameter, i.e., $\xi < 0.00064$ at 95\% CL.
We finally show that in this class of modified gravity models the use a prior on the absolute magnitude $M_B$ 
in combination with the Pantheon SNe sample leads to results which are very consistent with those obtained by 
imposing a prior on $H_0$, as happens for other {\em early-type} models which accommodate a 
larger value of $H_0$ compared to the $\Lambda$CDM results.
\end{abstract}

\maketitle

\section{INTRODUCTION}
The use of type Ia supernovae (SNe Ia), which can be calibrated fairly reliably to provide accurate 
distances as standard candles, can be used to map the expansion history of the Universe.

Although for the $\Lambda$CDM model SNe data have little statistical power compared to cosmic microwave 
background (CMB) anisotropies and baryon acoustic oscillations (BAO) measurements, they represent a 
precious tool to test extended models. The use of SNe Ia for cosmology has been studied extensively 
for many applications from the determination of the Hubble constant \cite{Branch:1992rv,Freedman:2010xv} 
to study models involving evolving dark energy 
\cite{SupernovaSearchTeam:1998fmf,SupernovaCosmologyProject:1998vns,Leibundgut:2001jd,Frieman:2008sn} 
and modified gravity \cite{Amendola:1999vu,Dainotti:2021pqg}.

SNe measurements can be used to determine the Hubble constant using the cosmic distance ladder calibration. 
The determination of the Hubble constant from Cepheid calibrated SNe Ia started to significantly depart 
from its inference within $\Lambda$CDM from CMB anisotropies since the first {\em Planck} data release 
(DR1) \cite{Planck:2013pxb}.
This discrepancy has now grown: the value inferred using CMB data from the {\em Planck} DR3 for a flat 
$\Lambda$CDM cosmological model, $H_0 = (67.36\pm 0.54)$ ${\rm km}\, {\rm s}^{-1}\, {\rm Mpc}^{-1}$ 
at 68\% confidence level (CL) \cite{Planck:2018vyg}, is in a $5.0\sigma$ tension\footnote{Quantified as 
$\left|H_0^{(1)}-H_0^{(2)}\right|/\sqrt{\sigma\left(H_0^{(1)}\right)^2+\sigma\left(H_0^{(2)}\right)^2}$.}
with the measurement from the SH0ES team \cite{Riess:2021jrx} obtained with cosmic distance ladder 
calibration of SNe Ia using the Cepheids measured in the host of 42 SNe Ia from the revised Pantheon+ 
compilation \cite{Carr:2021lcj}, that is $H_0 = (73.0\pm 1.0)$ ${\rm km}\, {\rm s}^{-1}\, {\rm Mpc}^{-1}$ 
at 68\% CL.
See Ref.~\cite{Riess:2021jrx} for an updated discussion of uncertainties in calibration or in the 
luminosity functions of SNe Ia which could affect the tension between the local determination of $H_0$ 
and its inference in $\Lambda$CDM from several cosmological measurements.

Cosmic distance ladder calibration does not directly extract the value of the the Hubble constant and 
it has been suggested to translate the Hubble constant tension into a SNe absolute magnitude tension 
rather than in $H_0$ \cite{Camarena:2019moy,Efstathiou:2021ocp,Camarena:2021jlr}. In this regard, late-time 
solutions to the Hubble tension have been questioned as valid solutions \cite{Efstathiou:2021ocp,Camarena:2021jlr}, 
since the estimated value of the Hubble constant from the SNe absolute magnitude depends from the late-time 
background expansion. Indeed, the use of a prior information on the intrinsic magnitude rather than on the Hubble 
constant as already been tested for some extensions of the $\Lambda$CDM cosmological model in 
Ref.~\cite{Schoneberg:2021qvd}. 
Many early-time extensions of the $\Lambda$CDM model have been proposed in order to tackle 
the Hubble tension, such as early dark energy \cite{Karwal:2016vyq,Poulin:2018cxd,Agrawal:2019lmo,Braglia:2020bym}, 
modified gravity \cite{Umilta:2015cta,Ballardini:2020iws,Braglia:2020auw}, 
primordial magnetic fields \cite{Jedamzik:2020krr}, 
and variation of fundamental constants \cite{Hart:2019dxi,Ballardini:2021evv}; 
see Refs.~\cite{Knox:2019rjx,DiValentino:2020zio,Jedamzik:2020zmd,DiValentino:2021izs,Perivolaropoulos:2021jda,Shah:2021onj,Abdalla:2022yfr} for reviews on the topic.

Modified gravity has an impact on the astrophysics of SNe Ia. By allowing the gravitational constant to vary 
with redshift, the dependence of the peak luminosity of the SNe from the mass of the white dwarf progenitors 
becomes redshift dependent \cite{Amendola:1999vu,GarciaBerro:1999bq}.
The correction due to the evolution of intrinsic luminosity to an evolution of the value of Newton’s 
gravitational constant $G$ has been studied and it has also been utilized to place constraints on the 
variation of $G$ using the observational dispersion in SNe Ia absolute magnitudes, see 
Refs.~\cite{Amendola:1999vu,GarciaBerro:1999bq,Gaztanaga:2001fh,Riazuelo:2001mg,Lazkoz:2005sp,Nesseris:2006jc,Finelli:2007wb,Mould:2014iga,Wright:2017rsu,Desmond:2019ygn,Sapone:2020wwz,Marra:2021fvf}.
This means that for accurate cosmological studies of gravity, we must carefully consider modified 
gravity’s impact on SNe Ia astrophysics and its implication in terms of cosmological parameter inference.

In this paper, we assess the impact of adding SNe data to cosmological analyses alone and in combination 
with {\em Planck} DR3 and BAO from Sloan Digital Sky Survey (SDSS) data to constrain as a working example 
one of the simplest scalar-tensor gravity model such as induced gravity (IG), equivalent to Jordan-Brans-Dicke 
\cite{Jordan:1949zz,Brans:1961sx}, described by the action 
\begin{equation} \label{eqn:action}
    S = \int \dd^{4}x \sqrt{-g} \left[ \xi\sigma^2\frac{R}{2} 
    - \frac{g^{\mu\nu}}{2} \partial_\mu \sigma \partial_\nu \sigma - V(\sigma) + {\cal L}_{\rm m} \right]
\end{equation}
where $\xi > 0$ is the coupling parameters, $\sigma$ is a scalar field, $R$ is the Ricci scalar, and 
${\cal L}_{\rm m}$ is the Lagrangian density for matter fields minimally coupled to the metric. We restrict 
ourselves to a potential of the type $V(\sigma) \propto \sigma^4$ \cite{Amendola:1999qq,Cooper:1981byv,Finelli:2007wb} 
in which the scalar field is effectively massless and the effective gravitational constant $G_{\rm eff}$ 
between two test masses is \cite{Boisseau:2000pr}
\begin{equation} \label{eqn:boundary}
    G_{\rm eff}(z=0) = \frac{1}{8 \pi \xi\sigma_0^2}\frac{1 + 8\xi}{1 + 6\xi} \frac{1}{\left(1+\Delta\right)^2} \,.
\end{equation}
Following Ref.~\cite{Ballardini:2021evv} (see also Ref.~\cite{Joudaki:2020shz}), we 
introduce an imbalance $\Delta$ between the gravitational constant today $G_{\rm eff}(z=0)$ and 
the Newton constant $G$.
Note that, scalar-tensor theories of gravity that involve a scalar field nonminimally coupled to the Ricci 
scalar naturally lead to a higher CMB-inferred value of $H_0$ 
\cite{Umilta:2015cta,Ballardini:2016cvy,Rossi:2019lgt,Braglia:2020iik,Ballardini:2020iws,Braglia:2020auw,Joudaki:2020shz,Abadi:2020hbr,Ballardini:2021evv} and can also help in interpreting the current tensions in the estimates of cosmological parameters from different observations.
With the working example adopted here we can therefore also test the difference between
a prior on the absolute magnitude and on $H_0$.\footnote{Note also that a parametric sudden transition of the effective gravitational constant at late 
times \cite{Marra:2021fvf,Alestas:2021luu} could also reduce the tension in the Hubble constant 
(see Refs.~\cite{DiValentino:2020zio,Schoneberg:2021qvd} 
for a review and comparison of models able to alleviate the $H_0$ tension).}

Our paper is organized as follows. After this introduction, we describe the distance modulus relation 
used to derive constraints from SNe Ia data including the correction due to the evolution of the 
Newton's constant and we quantify the impact on current cosmological data in Sec.~\ref{sec:Chandrasekhar}. 
In Sec.~\ref{sec:results}, we describe the datasets and prior considered and we present our results 
in light of SNe Ia data alone and in combination with CMB and BAO data. We study the impact of using a 
prior on the absolute magnitude, based on the SH0ES calibration, together to the SNe Ia sample in 
Sec.~\ref{sec:MB_prior}. In Sec.~\ref{sec:conclusion} we draw our conclusions.
In the Appendix, we assess the importance of including the correct redshift dependence of the 
Chandrasekhar mass in the SNe likelihood for the models considered.

\section{THE DISTANCE MODULUS RELATION IN SCALAR-TENSOR THEORIES} \label{sec:Chandrasekhar}
The peculiarity of SNe Ia is their nearly uniform intrinsic luminosity with an absolute magnitude 
around $M \sim -19.5$ \cite{Sahni:1999gb,Carroll:2000fy}, and this allows us to promote SNe Ia to a 
well-established class of standard candles. To evaluate the underlying best cosmological model, we make 
use of the distance modulus $\mu$, derived from the observations of SNe Ia, and we compare it with the 
theoretical one $\mu_{\rm th}$, defined as follows
\begin{align} \label{eqn:mu_th}
    \mu_{\rm th} &= m_B - M_B \\
                 &= 5 \log_{10} d_L(z) + 25 \quad [{\rm mag}]
\end{align}
where $m_B$ is the $B$-band apparent magnitude of the source, $M_B$ is the absolute magnitude in the $B$-band, 
and $d_L$ is the luminosity distance defined as \cite{Kenworthy:2019qwq}
\begin{equation}
    d_L(z) = c(1+z_{\rm hel})
    \int_0^{z_{\rm cmb}}\frac{{\rm d}z'}{H(z')}  \quad [{\rm Mpc}]
\end{equation}
where $z_{\rm hel}$ is the heliocentric redshift and $z_{\rm cmb}$ is the CMB redshift corrected by 
peculiar velocities. Finally, the observed distance modulus $\mu_{\rm obs}$ is defined as
\begin{equation} \label{eqn:mu_obs}
    \mu_{\rm obs} = m_B - M_B + \alpha x_1 - \beta c + \Delta M + \Delta B \quad [{\rm mag}]
\end{equation}
where $\alpha$ is the coefficient of the relation between luminosity and stretch, $x_1$ is the stretch 
parameter, $\beta$ is the coefficient of the relation between luminosity and the color, $c$ is the color, 
$\Delta M$ is a distance correction based on the host-galaxy mass of the SN, and $\Delta B$ is a bias 
correction based on simulations.
The difficulty in the determination of the cosmological parameters lies in the identification of $M_B$, 
as detailed in \cite{Scolnic:2017caz}.

The evolution of Newton constant predicted in the context of modified gravity theories induces 
special effects to the physics of SNe Ia; see 
Refs.~\cite{Amendola:1999vu,GarciaBerro:1999bq,Gaztanaga:2001fh,Riazuelo:2001mg,Lazkoz:2005sp,Nesseris:2006jc,Finelli:2007wb,Wright:2017rsu,Desmond:2019ygn,Sapone:2020wwz,Marra:2021fvf}.
The observed magnitude redshift relation of SNe Ia can be translated to luminosity distance-redshift 
relation [which leads to the expansion history $H(z)$] only under the assumption that SNe Ia behave as 
standard candles, in particular the constancy in time of the Chandrasekhar mass.
The peak luminosity of SNe Ia is proportional to the mass of nickel synthesized which is a fixed fraction 
of the Chandrasekhar mass $M_{\rm Ch} \sim G^{-3/2}$; see Refs.~\cite{Gaztanaga:2001fh,Wright:2017rsu}. 
Therefore the SN Ia peak luminosity varies like $L \sim G^{-3/2}$ \cite{Arnett:1982ioj} and the 
corresponding distance modulus \eqref{eqn:mu_th} in presence of a varying effective gravitational constant becomes 
\begin{equation} \label{eqn:mu_Ch}
    \mu_{\rm th}(z) = 5 \log_{10} d_L(z) + 25 + \frac{15}{4} \log_{10} \frac{G_{\rm eff}(z)}{G}  \quad [{\rm mag}] \,.
    \end{equation}
Since the absolute magnitude $M_B$ is  marginalized in Eq.~\eqref{eqn:mu_obs} as being a nuisance 
parameter, its possible redshift dependence may carry useful information about the robustness of the 
determination of $H_0$ using SNe Ia data and about possible modifications of $G_{\rm eff}(z)$. 

Note also that there are modified gravity models in which deviations of $G_{\rm eff}(z)$ from $G$ at late times 
are so small by construction that the correction due to the redshift dependence of the Chandrasekhar mass is 
expected to be negligible \cite{Braglia:2020auw}.

\section{CONSTRAINTS FROM COSMOLOGICAL OBSERVATIONS IN COMBINATION WITH TYPE IA SUPERNOVAE} \label{sec:results}
In this section, we present the constraints on cosmological parameters including Pantheon SNe to the 
results presented in Ref.~\cite{Ballardini:2021evv} obtained from the combination of {\em Planck} 
2018 DR3 (hereafter P18) with BAO measurements from BOSS DR12.
In addition, we show the constraints on the modify gravity parameters from the compilation of SNe alone.

We consider the full CMB information from {\em Planck} DR3 \cite{Aghanim:2019ame,Aghanim:2018oex} 
including the low-$\ell$ likelihood {\tt Commander} (temperature-only) plus the {\tt SimAll} (EE-only), 
the high-multipoles likelihood {\tt Plik}, and the CMB lensing likelihood on the {\em conservative} 
multipoles range, i.e., $8 \leq \ell \leq 400$. 
The compilation of BAO data includes data from Baryon Spectroscopic Survey (BOSS) DR12 
\cite{Alam:2016hwk} {\em consensus} results in three redshift slices with effective redshifts 
$z_{\rm eff} = 0.38,\,0.51,\,0.61$, the measure from 6dF \cite{Beutler:2011hx} at $z_{\rm eff} = 0.106$, 
and the one from SDSS DR7 \cite{Ross:2014qpa} at $z_{\rm eff} = 0.15$.
We consider the Pantheon sample which is a compilation of 1048 spectroscopically confirmed SNe Ia that 
gathers different surveys spanning the redshift range $0.01 < z < 2.3$\footnote{\href{https://github.com/dscolnic/Pantheon}{https://github.com/dscolnic/Pantheon}} \cite{Scolnic:2017caz}.

We use {\tt MontePython}\footnote{\href{https://github.com/brinckmann/montepython\_public}{https://github.com/brinckmann/montepython\_public}} \cite{Audren:2012wb,Brinckmann:2018cvx} connected to our modified 
version of the code {\tt CLASS}\footnote{\href{https://github.com/lesgourg/class\_public}{https://github.com/lesgourg/class\_public}} \cite{Lesgourgues:2011re,Blas:2011rf}, i.e., {\tt CLASSig} \cite{Umilta:2015cta}. 
For the Markov chain Monte Carlo (MCMC) analysis including CMB data (P18, P18 + SNe, P18 + BAO, P18 + BAO + SNe), we vary the six 
cosmological parameters for a flat $\Lambda$CDM concordance model, i.e., $\omega_{\rm b}$, $\omega_{\rm c}$, $H_0$, 
$\tau$, $\ln\left(10^{10}A_{\rm s}\right)$, and $n_{\rm s}$; for the analysis of SNe compilation alone we vary 
$\Omega_{\rm cdm}$ fixing $\Omega_{\rm b} = 0.0047$.
The extra parameters related to the coupling to the Ricci curvature are 
$\zeta_{\rm IG} \equiv \ln\left(1 + 4\xi\right)$, sampled in the prior range $[0,\,0.039]$ and 
$\Delta \in [-0.3,\, 0.3]$. We assume two massless neutrino with $N_{\rm eff} = 2.0328$, and a massive 
one with fixed minimum mass $m_\nu = 0.06$ eV. 
We assume adiabatic initial condition for the scalar fluctuations \cite{Paoletti:2018xet}.  
We set the primordial helium abundance according to the prediction from {\tt PArthENoPE} \cite{Pisanti:2007hk,Consiglio:2017pot} taking into account the effect of a different gravitational constant 
as a source of extra radiation in $Y_{\rm BBN}(\omega_{\rm b},\,N_{\rm eff})$ 
\cite{Ballardini:2016cvy,Ballardini:2021evv}.
We vary also nuisance and foreground parameters for the {\em Planck} and Pantheon likelihoods.

\begin{figure*}
\centering
\includegraphics[width=0.9\textwidth]{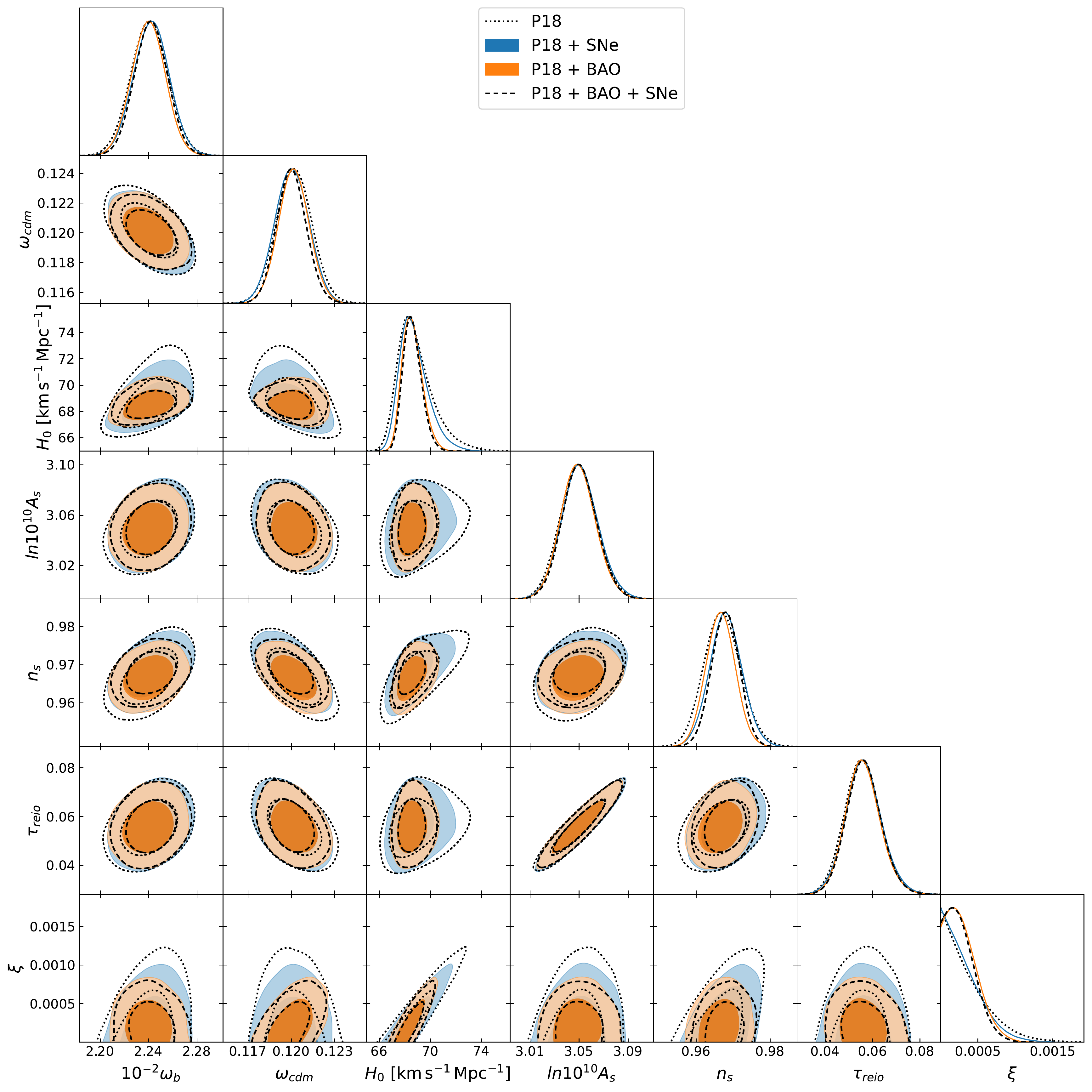}
\caption{Marginalized joint 68\% and 95\% CL regions 2D parameter space using the {\em Planck} 2018 DR3 
alone (blue contours) and in combination with BAO data (orange contours) for IG.
The dotted and dashed contours include the Pantheon SNe sample for P18 and P18 + BAO, respectively.}
\label{fig:SNe_ig}
\end{figure*}
In Fig.~\ref{fig:SNe_ig}, we compare the marginalized constraints on cosmological parameters for P18 and 
P18 + BAO including Pantheon SNe. The addition of SNe data slightly improves the constraints on cosmological 
parameters compared to the P18-only case and P18 + BAO combination, for IG we find
\begin{equation}
\xi < \begin{cases} 
0.0095 & \mbox{SNe} \\
0.00096 & \mbox{P18} \\
0.00080 & \mbox{P18 + SNe} \\
0.00068 & \mbox{P18 + BAO} \\
0.00063 & \mbox{P18 + BAO + SNe}
\end{cases}
\end{equation}
at 95\% CL for coupling parameter and for the Hubble constant at 68\% CL  
\begin{equation}
H_0 = \begin{cases} 
68.8^{+0.8}_{-1.7}\ {\rm km}\,{\rm s}^{-1}\,{\rm Mpc}^{-1} & \mbox{P18} \\
68.7^{+0.8}_{-1.3}\ {\rm km}\,{\rm s}^{-1}\,{\rm Mpc}^{-1} & \mbox{P18 + SNe} \\
68.6^{+0.6}_{-0.9}\ {\rm km}\,{\rm s}^{-1}\,{\rm Mpc}^{-1} & \mbox{P18 + BAO} \\
68.6^{+0.6}_{-0.8}\ {\rm km}\,{\rm s}^{-1}\,{\rm Mpc}^{-1} & \mbox{P18 + BAO + SNe} \,.
\end{cases}
\end{equation}

\begin{figure*}
\centering
\includegraphics[width=0.9\textwidth]{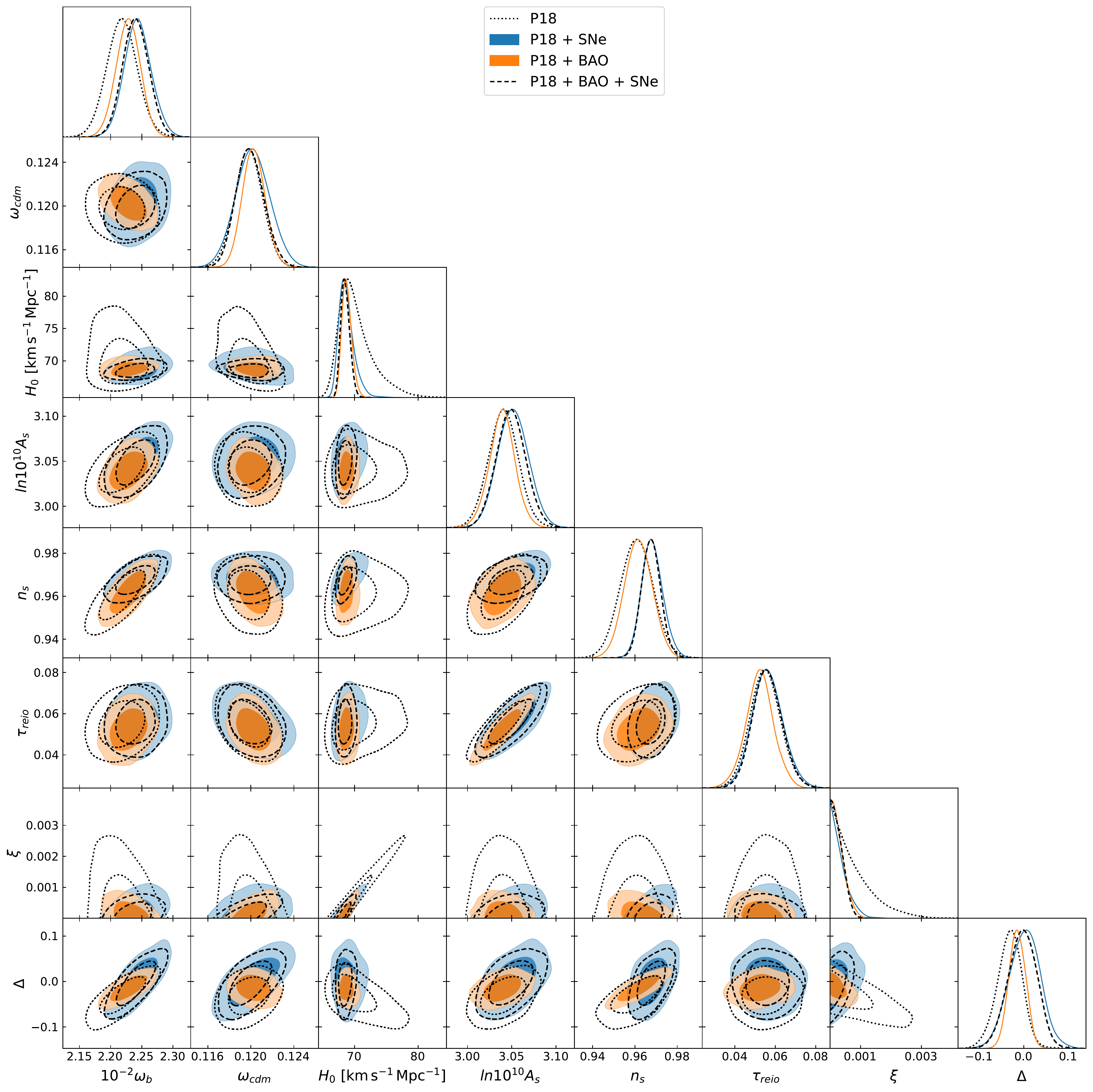}
\caption{As in Fig.~\ref{fig:SNe_ig}, for $\Delta$ allowed to vary.} 
\label{fig:SNe_dig}
\end{figure*}

Relaxing the consistency condition on the current value of the effective gravitational constant, i.e., 
Eq.~\eqref{eqn:boundary} with $\Delta$ allowed to vary, we find (see Fig.~\ref{fig:SNe_dig}) that the imbalance is constrained at 68\% CL to 
\begin{equation}
\Delta = \begin{cases} 
-0.032^{+0.029}_{-0.025} & \mbox{P18} \\
0.002^{+0.037}_{-0.032} & \mbox{P18 + SNe} \\
-0.022\pm 0.023 & \mbox{P18 + BAO} \\
-0.003^{+0.034}_{-0.030} & \mbox{P18 + BAO + SNe} 
\end{cases}
\end{equation}
\begin{equation}
\xi < \begin{cases} 
0.0096 & \mbox{SNe}\,, \\
0.0021 & \mbox{P18}\,, \\
0.00088 & \mbox{P18 + SNe} \\
0.00082 & \mbox{P18 + BAO} \\
0.00064 & \mbox{P18 + BAO + SNe} 
\end{cases}
\end{equation}
at 95\% CL for coupling parameter and for the Hubble constant at 68\% CL  
\begin{equation}
H_0 = \begin{cases} 
70.2^{+1.2}_{-3.1}\ {\rm km}\,{\rm s}^{-1}\,{\rm Mpc}^{-1} & \mbox{P18} \\
68.7^{+0.8}_{-1.4}\ {\rm km}\,{\rm s}^{-1}\,{\rm Mpc}^{-1} & \mbox{P18 + SNe} \\
68.6^{+0.7}_{-0.9}\ {\rm km}\,{\rm s}^{-1}\,{\rm Mpc}^{-1} & \mbox{P18 + BAO} \\
68.6^{+0.6}_{-0.8}\ {\rm km}\,{\rm s}^{-1}\,{\rm Mpc}^{-1} & \mbox{P18 + BAO + SNe} \,.
\end{cases}
\end{equation}
In this case, the addition of SNe data is appreciable leading to tighter constraints for both the standard 
parameters and the modified gravity ones; moreover the mean value of the imbalance is more consistent to 
the Newton constant, i.e., $\Delta = 0$. We find that using the SNe data alone we are not able to put 
constraints on $\Delta$ on the prior range assumed in the analysis, i.e., $[-0.3,\, 0.3]$.
It is interesting to note that the addition of SNe data to P18 and to P18 + BAO leads to larger uncertainties 
on $\Delta$. 
Reducing the degeneracy between the coupling parameter $\xi$ and the total matter density parameter 
$\Omega_{\rm m}$, the addition of SNe data help in constraining $\xi$. Consequently, the bound on 
the imbalance $\Delta$, which is not constrained by the SNe data alone, is relaxed due to the 
partial degeneracy with the coupling $\xi$.
We check the value of the $\Delta \chi^2$ with respect to the $\Lambda$CDM case, 
calculated as $\Delta \chi^2 = \chi^2 - \chi^2_{\rm \Lambda CDM}$. 
We find that for IG with $\Delta = 0$ ($\Delta$ allowed to vary) the $\Delta \chi^2$ 
corresponds to -1.4 (-2.6) for P18 + SNe, -2.2 (-3.6) for P18 + BAO, and -2.4 (-3.7) 
for P18 + BAO + SNe pointing to a slight improvement of fit for IG.
We also calculate the Bayes factor with respect to the $\Lambda$CDM case, 
calculated as the ratio of the evidences for the extended model with respect to the baseline $\Lambda$CDM model.  
We compute the evidence directly from our MCMC chains using the method introduced in Ref.~\cite{Heavens:2017afc}\footnote{\href{https://github.com/yabebalFantaye/MCEvidence}{https://github.com/yabebalFantaye/MCEvidence}}.
The logarithmic of the Bayes factor $\ln B$ for IG with $\Delta = 0$ ($\Delta$ allowed to vary) corresponds to 
-0.4 (-1.9) for P18 + SNe, -1.2 (-2.9) for P18 + BAO, and to -1.6 (-2.7) for P18 + BAO + SNe showing no statistical 
preference for the model analyzed \cite{Nesseris:2012cq}. 
Note that, the Bayes factor depends on the prior range of the parameters and 
it is enhanced in presence of parameters not well constrained.
We interpret the $1\sigma$ shift of the mean value of $\Delta$, coming with not significant effects on the 
$\Delta \chi^2$ and $\ln B$, as a sign that this parameter is weakly constrained 
from the datasets considered and it can moved along its degeneracy with $\xi$.

Finally we can project the constraints on $\Delta$ on the value of the Newton constant 
which is constrained at 68\% CL to 
$G_{\rm eff}(z=0)/G = 0.938^{+0.056}_{-0.049}$ for P18, 
$G_{\rm eff}(z=0)/G = 1.005\pm 0.071$ for P18 + SNe, 
$G_{\rm eff}(z=0)/G = 0.957\pm 0.045$ for P18 + BAO, and 
$G_{\rm eff}(z=0)/G = 1.05^{+0.05}_{-0.08}$ for P18 + BAO + SNe.

\section{THE ABSOLUTE MAGNITUDE AND THE HUBBLE PARAMETER PRIORS} \label{sec:MB_prior}
As explained in Refs.~\cite{Tripp:1997wt,Scolnic:2017caz}, there is a degeneracy between $H_0$ and $M$ 
fitting the distance modulus to a SN sample. 
For this reason, as pointed out in Refs.~\cite{Camarena:2021jlr,Efstathiou:2021ocp}, it is useful to 
look at corresponding constraints on the absolute magnitude $M_B$ of Pantheon SNe Ia sample rather 
than imposed the $H_0$ prior from SH0ES on the Hubble parameter at $z = 0$ in order to avoid misleading 
findings for late-time $\Lambda$CDM modifications as shown in 
Refs.~\cite{Lemos:2018smw,Camarena:2019moy,Camarena:2021jlr,Marra:2021fvf}. 
Indeed, the SH0ES Cepheid photometry \cite{Reid:2019tiq,Riess:2020fzl} and Pantheon SNe peak magnitudes 
give 
\begin{equation}
    M_B = -19.2435 \pm 0.0373 \quad [{\rm mag}] \,.
\end{equation}
This corresponds to $H_0 = 73.2 \pm 1.3\ {\rm km}\,{\rm s}^{-1} {\rm Mpc}^{-1}$ by fitting the Pantheon sample 
\cite{Scolnic:2017caz} with the low-redshift expansion to the luminosity distance in a $\Lambda$CDM background 
\cite{Camarena:2021jlr}.

\begin{figure}
\centering
\includegraphics[width=0.4\textwidth]{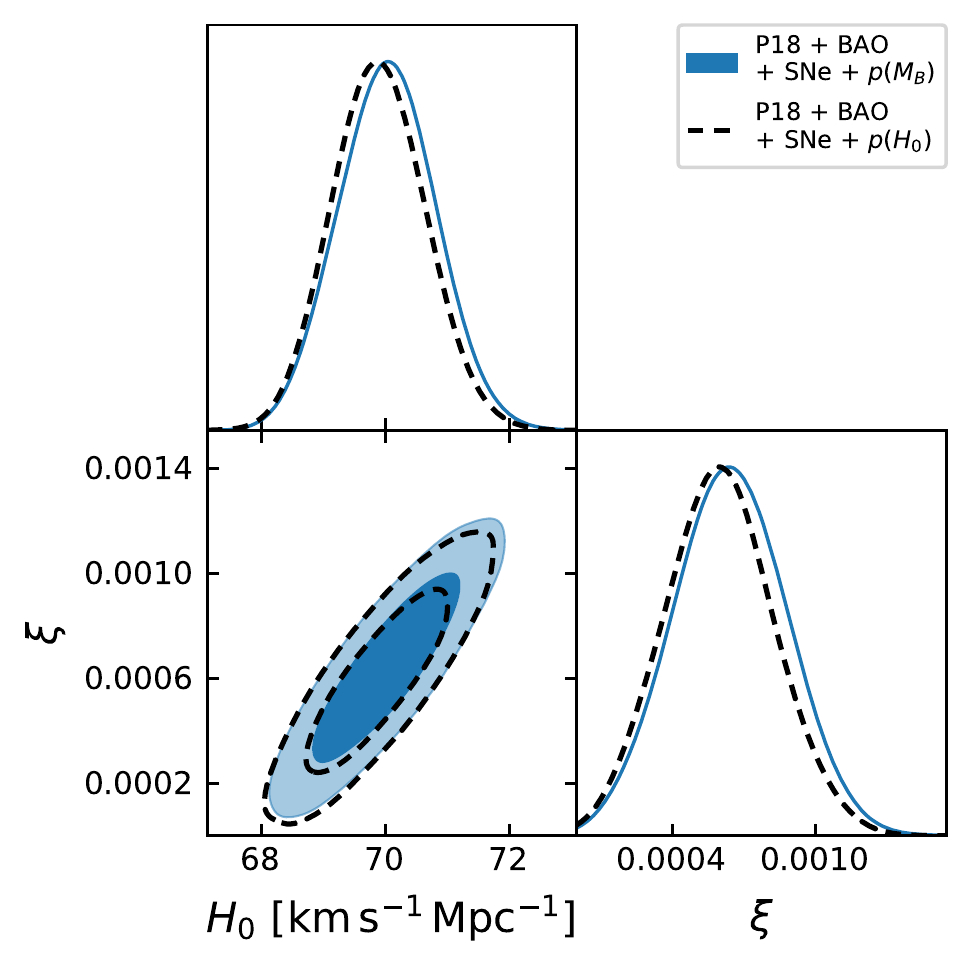}
\includegraphics[width=0.4\textwidth]{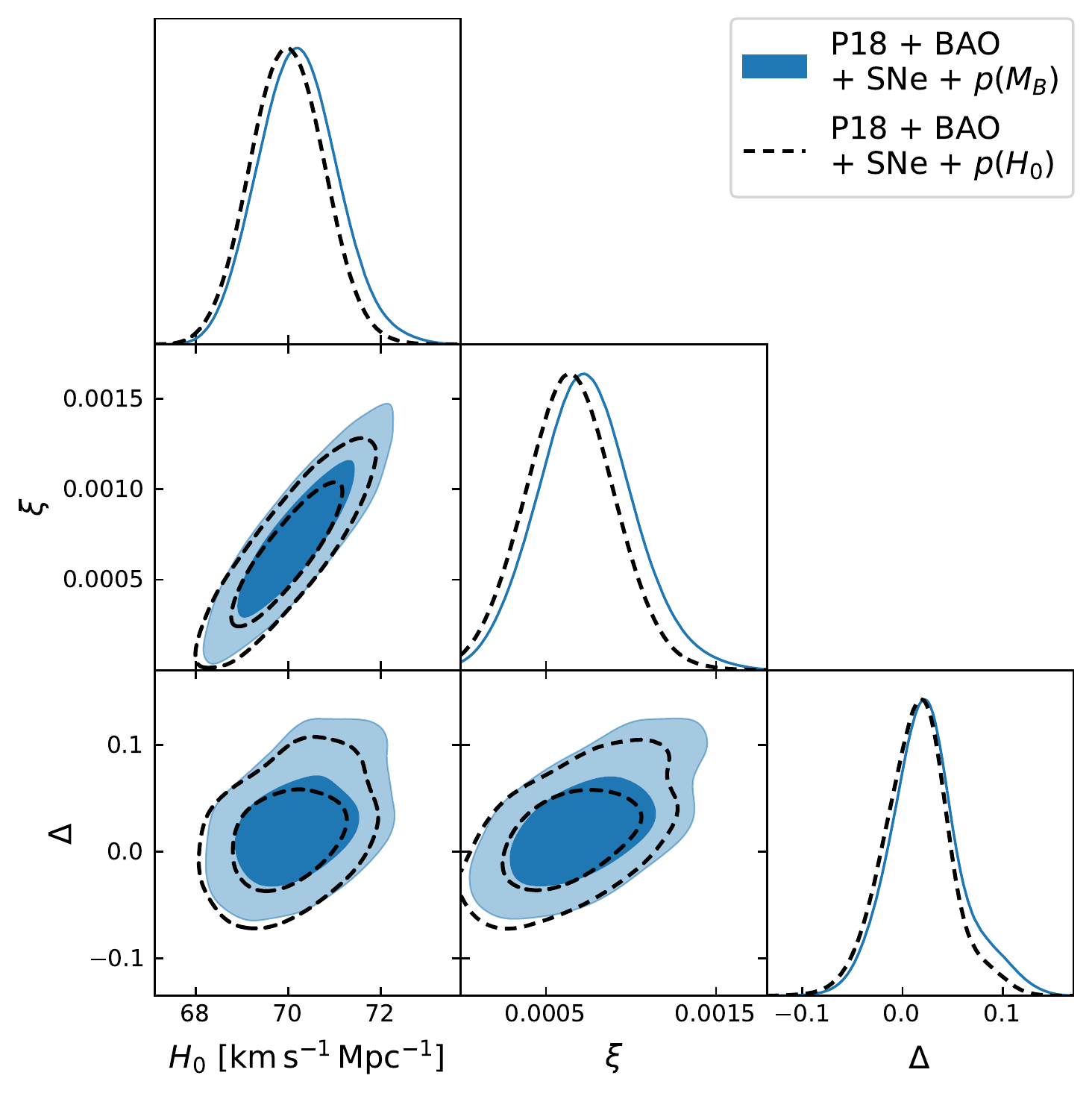}
\caption{Marginalized joint 68\% and 95\% CL regions 2D parameter space using the {\em Planck} 2018 
DR3 in combination with BAO data and Pantheon SNe sample, for IG with $\Delta = 0$ (top panel) and IG 
with $\Delta$ allowed to vary (bottom panel); the blue contours include a Gaussian prior on the absolute magnitude 
while dashed contours include a Gaussian prior on the Hubble constant.} \label{fig:SNe_ST_H0_bao}
\end{figure}

In Fig.~\ref{fig:SNe_ST_H0_bao}, we compare the marginalized constraints on cosmological parameters 
for P18 + BAO + SNe including the SH0ES information \cite{Riess:2020fzl} as a Gaussian prior on the 
Hubble parameter $p\left(H_0\right)$, corresponding to 
$H_0 = 73.2 \pm 1.3\, {\rm km}\, {\rm s}^{-1}\, {\rm Mpc}^{-1}$, versus a Gaussian prior on the absolute 
magnitude $p\left(M_B\right)$, corresponding to $M_B = -19.2435 \pm 0.0370\, {\rm mag}$ \cite{Camarena:2021jlr}.
\begin{figure}
\centering
\includegraphics[width=0.4\textwidth]{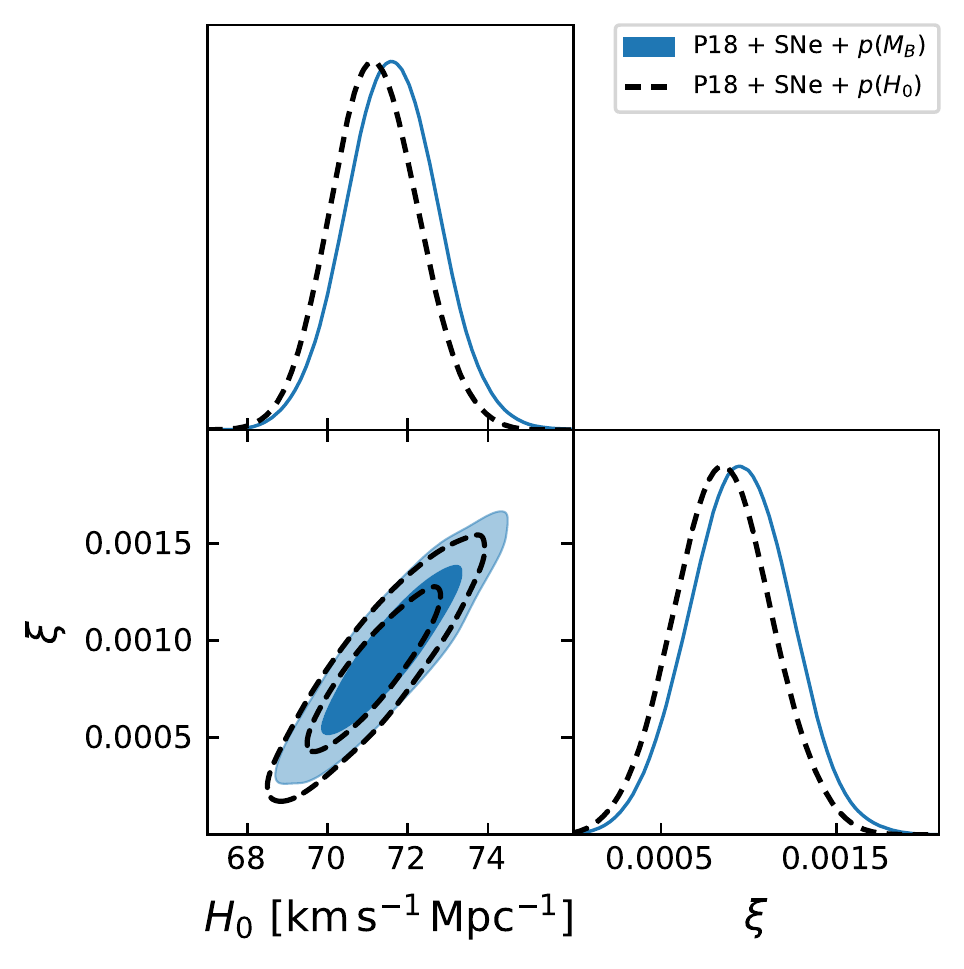}
\includegraphics[width=0.4\textwidth]{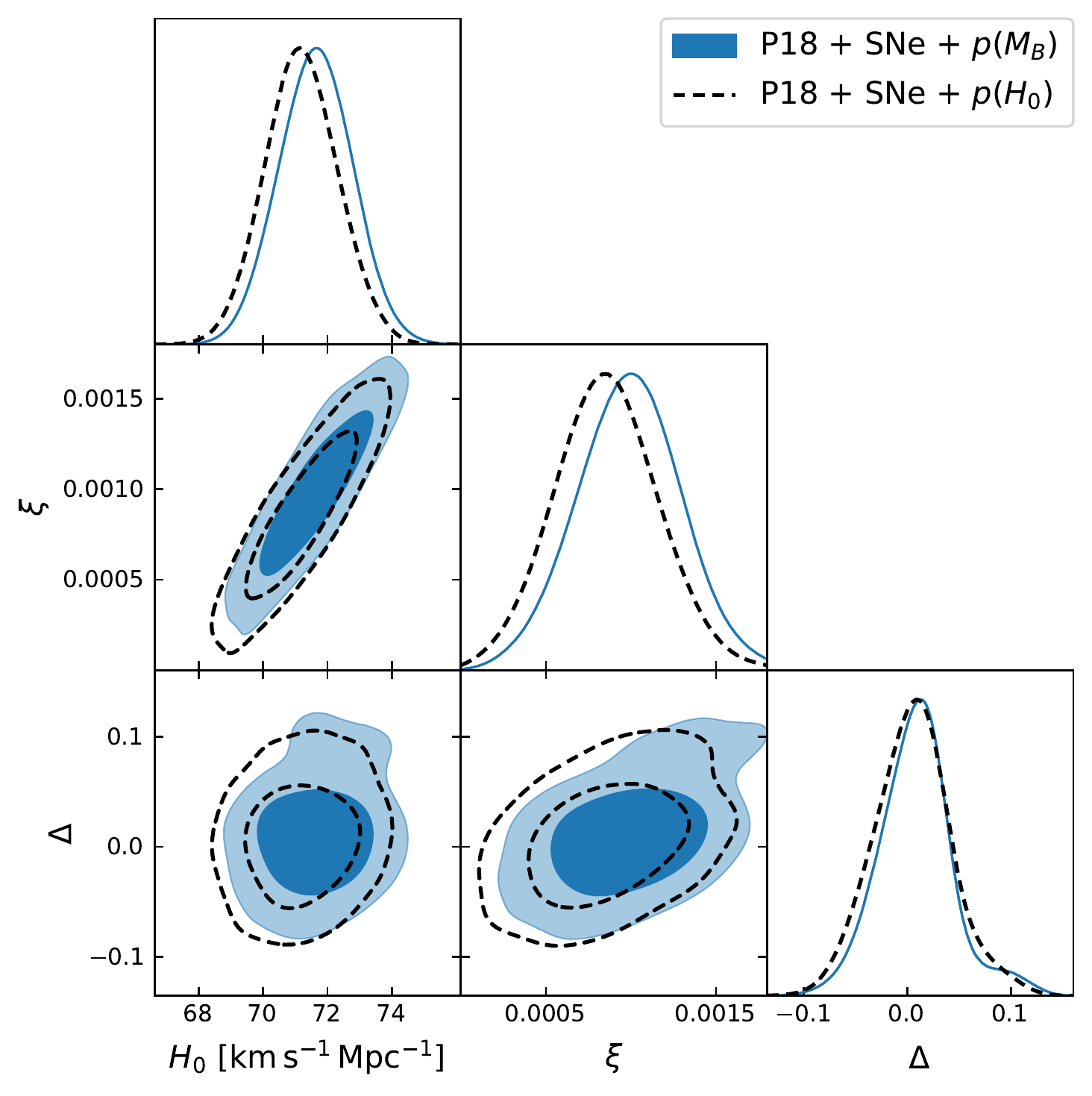}
\caption{As in Fig.~\ref{fig:SNe_ST_H0_bao}, without BAO data, i.e., P18 + SNe.}
\label{fig:SNe_ST_H0_nobao}
\end{figure}
We conclude that the use of a Gaussian prior on $H_0$ derived from the inverse distance ladder calibration 
of SNe, as done in previous studies of these scalar-tensor models in 
Refs.~\cite{Umilta:2015cta,Ballardini:2016cvy,Rossi:2019lgt,Braglia:2020iik,Ballardini:2020iws,Braglia:2020auw,Ballardini:2021evv}, is fully consistent with the most correct way of using a prior information on the absolute 
magnitude $M_B$. This result is robust when we allow for $G_{\rm eff} (z=0) \ne G$. 
Indeed, we find $M_B = -19.357 \pm 0.021$ ($M_B = -19.347 \pm 0.024$) adding a Gaussian prior 
on $H_0$ to the combination of P18 + SNe + BAO and $M_B = -19.353 \pm 0.022$ ($M_B = -19.354 \pm 0.023$) 
when we add a Gaussian prior on $M_B$ for IG with $\Delta = 0$ ($\Delta$ allowed to vary). Without including any 
extra information, we find $M_B = -19.393 \pm 0.020$ at 68\% CL for the same P18 + SNe + BAO combination of 
datasets for IG in both cases fixing $\Delta = 0$ or varying it.

We find for IG $H_0 = 71.6\pm 1.1\ {\rm km}\,{\rm s}^{-1}\,{\rm Mpc}^{-1}$ at 68\% CL for 
P18 + SNe + $p\left(M_B\right)$ for both $\Delta = 0$ and $\Delta$ allowed to vary; see Fig.~\ref{fig:SNe_ST_H0_nobao}.
For IG with $\Delta = 0$ ($\Delta$ allowed to vary), we find that the tension goes from $2.9\sigma$ ($2.9\sigma$) to 
$0.9\sigma$ ($0.9\sigma$) including the prior on the absolute magnitude. With the addition of BAO, we find for 
IG with $\Delta = 0$ $H_0 = 70.2\pm 0.9\ {\rm km}\,{\rm s}^{-1}\,{\rm Mpc}^{-1}$ and for $\Delta$ allowed to vary  
$H_0 = 70.0\pm 0.8\ {\rm km}\,{\rm s}^{-1}\,{\rm Mpc}^{-1}$ both at 68\% CL for 
P18 + SNe + BAO + $p\left(M_B\right)$. For IG with $\Delta = 0$ ($\Delta$ allowed to vary), we find that the tension goes 
from $3.2\sigma$ ($3.2\sigma$) to $1.9\sigma$ ($2.1\sigma$) including the prior on the absolute magnitude to 
the combination P18 + BAO + SNe. We have reported this last result for completeness although the tension between 
$H_0$ from P18 + SNe +BAO and from SH0ES is superior to 3$\sigma$.\footnote{Here the tension in terms of number 
of $\sigma$ has been calculated on the Hubble parameter with respect to the reference measure $H_0 = 73.2\pm1.3\ {\rm km}\,{\rm s}^{-1}\,{\rm Mpc}^{-1}$ \cite{Riess:2020fzl}.}

The consistency in using a Gaussian prior on $H_0$ or on $M_B$ is connected to 
smooth modification of the background expansion at late time in these models; see~\cite{Camarena:2021jlr}.

\section{CONCLUSION} \label{sec:conclusion}
The near future of cosmology will be fueled by a huge amount of data that will allow us to extract 
precise and accurate cosmological information.
The capability to test extended cosmological models beyond the minimal $\Lambda$CDM will 
be possible thanks to the combination of different datasets that will allow one 
to reduce degeneracies and to shrink parameter uncertainties by testing the evolution of Universe at different 
redshifts and scales.
Thus, a continuous progress in the modeling of cosmological observables for extended models is required, 
in order to avoid confusing systematics with new physics and avoid parameter bias.

In this paper, we explore the use and importance of SNe data for cosmological parameter inference 
in modified gravity settings in which the gravitational constant vary with redshift,
by taking as a working example induced gravity (IG), equivalent to Jordan-Brans-Dicke by a field redefinition.
Particularly, we study general constraints on the coupling parameter from the Pantheon compilation of 
type SNe Ia alone and in combination with {\em Planck} DR3 CMB data and BAO measurements from BOSS. 
We constrain $\xi < 0.0095$ (95\% CL) with Pantheon data, $\xi < 0.00080$ (95\% CL) for {\em Planck} 
in combination with Pantheon, and $\xi < 0.00063$ (95\% CL) from the combination of {\em Planck}, BOSS, 
and Pantheon.
Allowing also the imbalance $\Delta$ to vary, connected to an effective gravitational constant today 
different from the value of the bare gravitational constant $G_{\rm eff} = G(1+\Delta)^2$, the bound 
on the coupling parameter is slightly relaxed to $\xi < 0.0096$ (95\% CL) for Pantheon data, 
$\xi < 0.00088$ (95\% CL) for {\em Planck} in combination with Pantheon, and $\xi < 0.00064$ (95\% CL) 
for the combination of {\em Planck}, BOSS, and Pantheon.
For the imbalance, we find $\Delta = 0.002^{+0.037}_{-0.032}$ (68\% CL) for {\em Planck} in combination 
with Pantheon and $\Delta = -0.003^{+0.034}_{-0.030}$ (68\% CL) for the combination of {\em Planck}, 
BOSS, and Pantheon; Pantheon data alone cannot constrain $\Delta$.
In the Appendix~\ref{sec:app}, we show that the correction due to the redshift dependence of the Chandrasekhar 
mass in these models is small, but appreciable, when SN Ia are added to the combination P18 + BAO. 

We also test the use of a prior on the absolute magnitude $M_B$ instead of a prior on the Hubble 
constant $H_0$ from SH0ES observations for this class of modified gravity models finding that the results 
do not depend on the choice of prior information. 
Considering the combination P18 + BAO + SNe + $p\left(M_B\right)$, for IG we find 
$H_0 = 70.2\pm 0.9\ {\rm km}\,{\rm s}^{-1}\,{\rm Mpc}^{-1}$ (68\% CL), while 
$H_0 = 70.0\pm 0.8\ {\rm km}\,{\rm s}^{-1}\,{\rm Mpc}^{-1}$ (68\% CL) when allowing for the imbalance 
$\Delta$ to vary.
The robustness to different prior assumptions in $H_0$ or $M_B$ is consistent with the fact that the models 
studied here alleviate the tension in $H_0$ at early times.

\section*{Acknowledgments}
MB thanks Thejs Brinckmann for useful discussions on the Pantheon likelihood. 
MB and FF acknowledges financial support from the contract ASI/ INAF for the Euclid mission n.2018-23-HH.0. 
FF acknowledges also financial support by the agreement n. 2020-9-HH.0 ASI-UniRM2 ``Partecipazione italiana 
alla fase A della missione LiteBIRD".

\appendix
\section{Comparison of constraints with uncorrected distance modulus relation} \label{sec:app}
\begin{figure}
\centering
\includegraphics[width=0.48\textwidth]{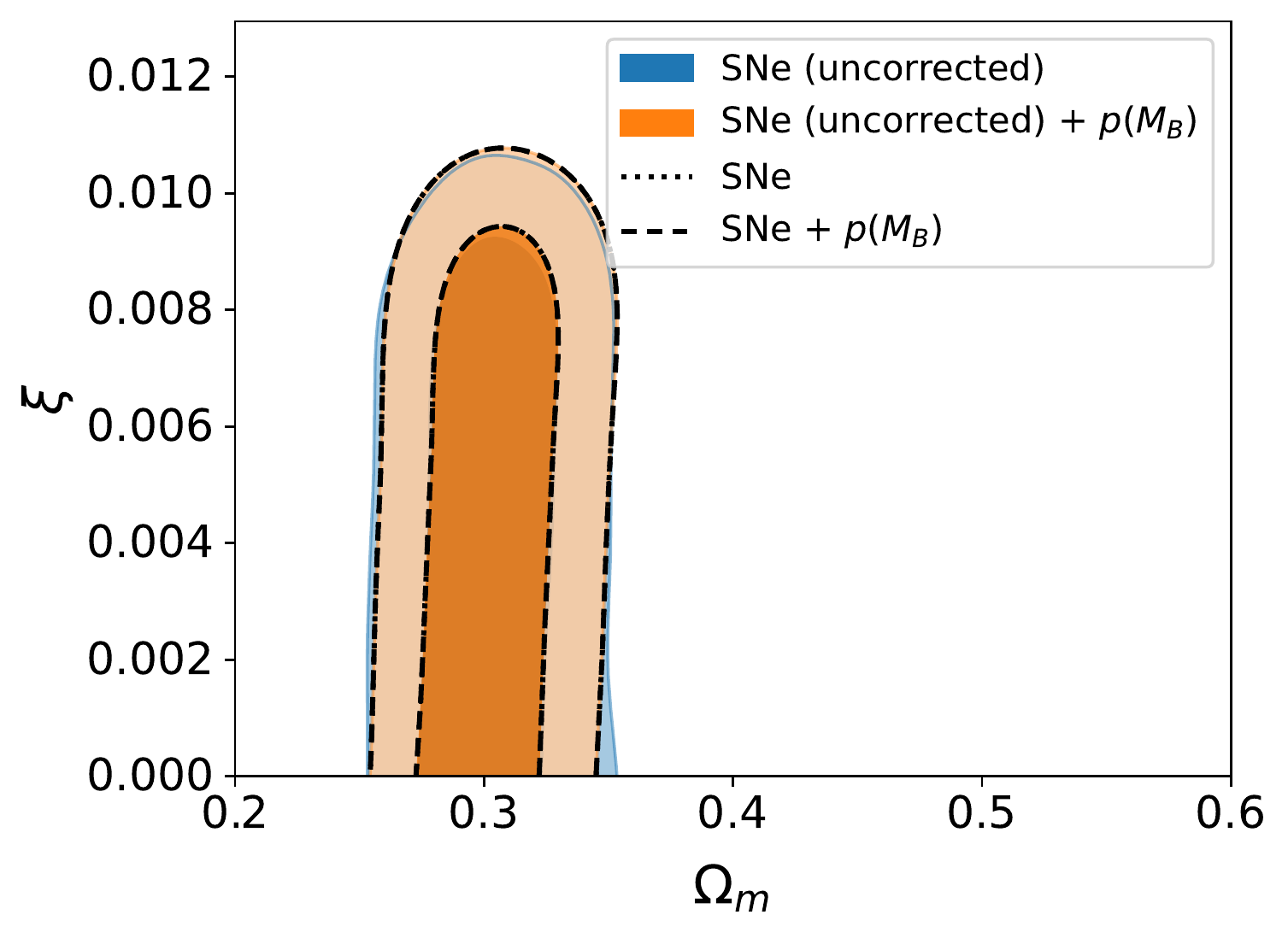}
\caption{Marginalized joint 68\% and 95\% CL regions 2D parameter space using the Pantheon SNe sample 
for IG (blue contours); orange contours include also a Gaussian prior on the absolute magnitude $p\left(M_B\right)$.
The dotted and dashed contours include the correction included in Eq.~\eqref{eqn:mu_Ch} for 
SNe and SNe + $p\left(M_B\right)$, respectively.}
\label{fig:SNe_Ch_ig}
\end{figure}

\begin{figure}
\centering
\includegraphics[width=0.48\textwidth]{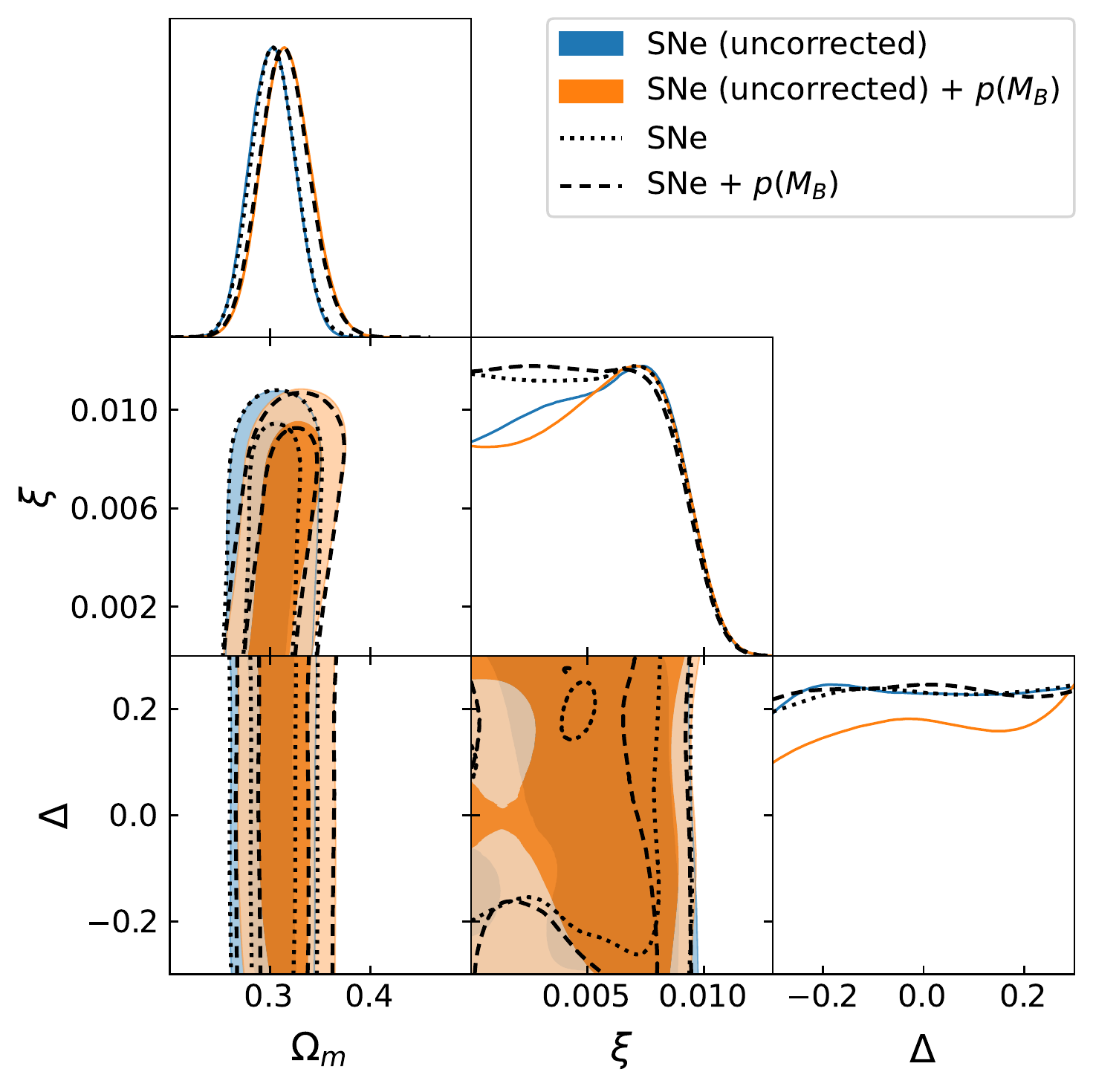}
\caption{As in Fig.~\ref{fig:SNe_Ch_ig}, for IG with $\Delta$ allowed to vary.}
\label{fig:SNe_Ch_dig}
\end{figure}

We test here the relevance of the redshift dependence of the Chandrasekhar mass in the SNe distance modulus likelihood. 
In Figs.~\ref{fig:SNe_Ch_ig} and \ref{fig:SNe_Ch_dig}, we compare the marginalized constraints using the SNe compilation alone with and 
without the correction to the distance modulus relation introduced in  Eq.~\eqref{eqn:mu_Ch}. 
When we analyze the SNe data alone, we only allowed $\Omega_{\rm m}$ to vary (fixing $\Omega_{\rm b} = 0.047$) 
in addition to IG parameters, instead we vary $\Omega_{\rm m}$ and $H_0$ analyzing the SNe data in combination to 
the prior on the absolute magnitude $p(M_B)$.
For IG with $\Delta = 0$, we find $\xi < 0.0093$ (95\% CL) and $\Omega_{\rm m} = 0.303\pm 0.023$ (68\% CL) 
without correction while we find $\xi < 0.0095$ (95\% CL) and $\Omega_{\rm m} = 0.320^{+0.021}_{-0.029}$ 
(68\% CL) correcting for Eq.~\eqref{eqn:mu_Ch}.
For IG with $\Delta$ allowed to vary, we find $\xi < 0.0095$ (95\% CL) and $\Omega_{\rm m} = 0.302\pm 0.022$ 
(68\% CL) without correction while we find $\xi < 0.0096$ (95\% CL) and $\Omega_{\rm m} = 0.317\pm 0.024$ 
(68\% CL) correcting for Eq.~\eqref{eqn:mu_Ch}. When analysing SNe data alone, we find a 
$0.5\sigma$ shift in the determination of the mean value of $\Omega_{\rm m}$ neglecting the redshift dependence of 
the Chandrasekhar mass.

\begin{figure*}
\centering
\includegraphics[width=0.98\textwidth]{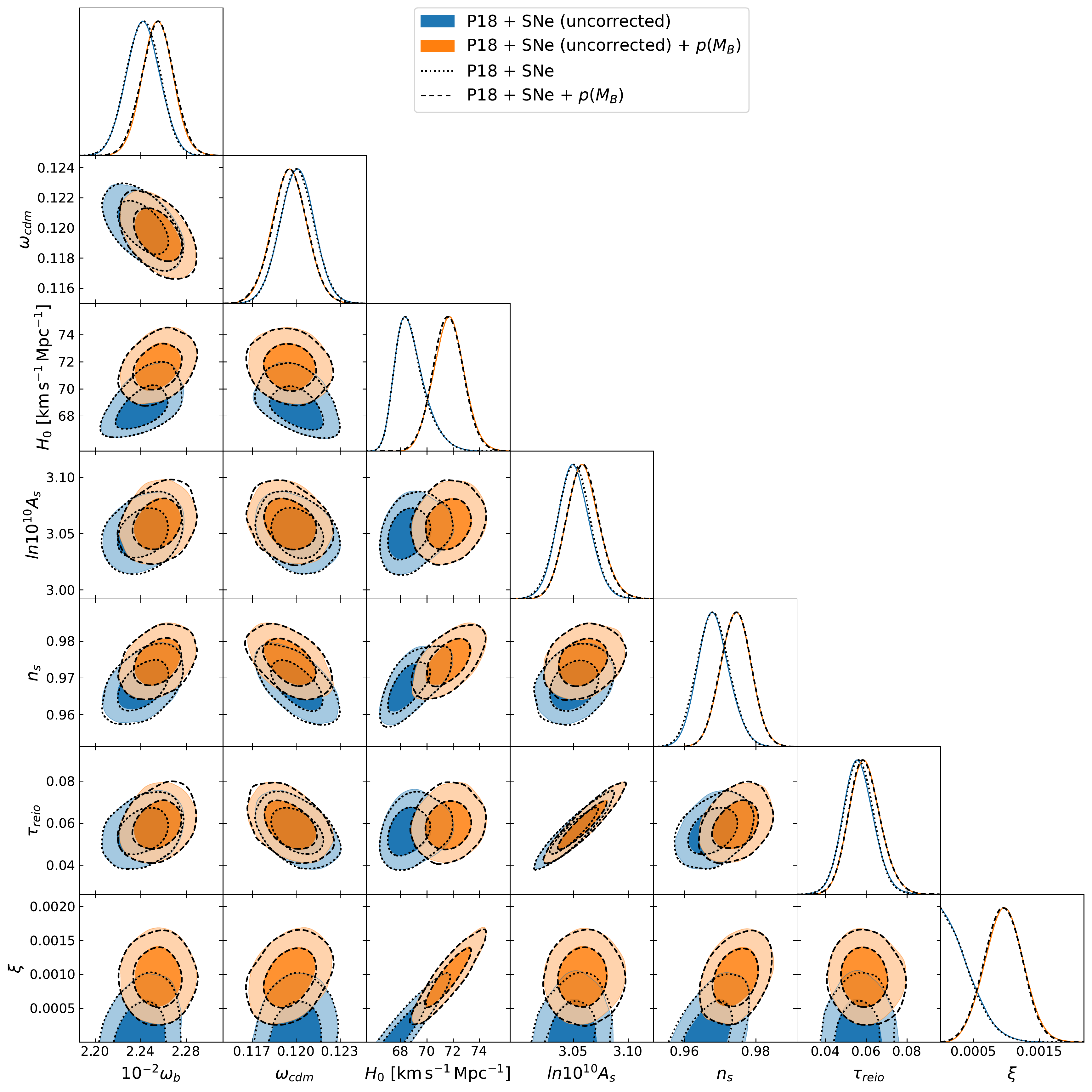}
\caption{Marginalized joint 68\% and 95\% CL regions 2D parameter space using the {\em Planck} 2018 
DR3 in combination with Pantheon SNe sample (blue contours) for IG; orange 
contours include also a Gaussian prior on the absolute magnitude $p\left(M_B\right)$.
The dotted and dashed contours include the correction included in Eq.~\eqref{eqn:mu_Ch} for 
P18 + SNe and P18 + SNe + $p\left(M_B\right)$, respectively.}
\label{fig:SNe_ig_Ch_nobao}
\end{figure*}

\begin{figure*}
\centering
\includegraphics[width=0.98\textwidth]{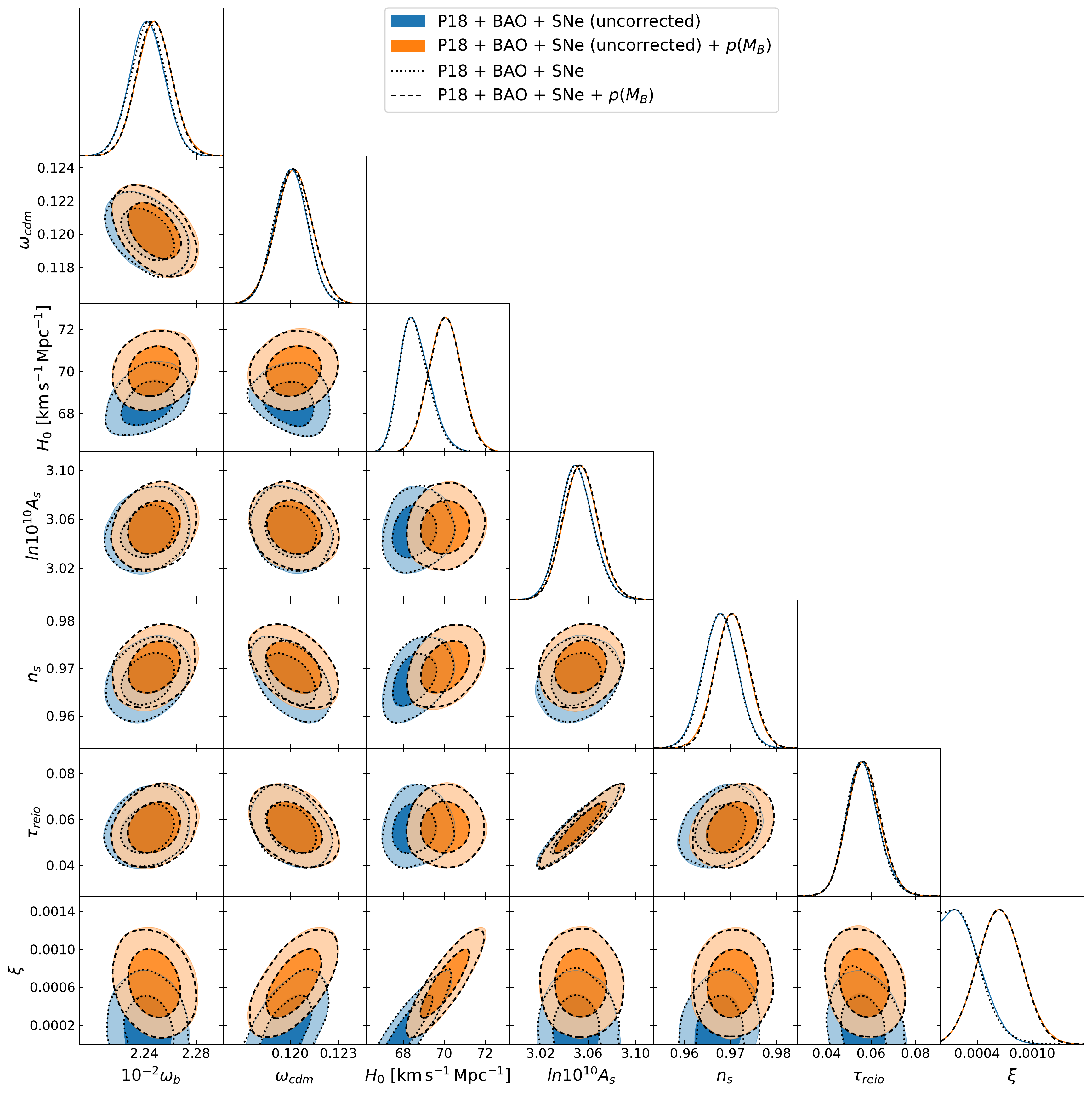}
\caption{Marginalized joint 68\% and 95\% CL regions 2D parameter space using the {\em Planck} 2018 
DR3 in combination with BAO data from BOSS DR12 and Pantheon SNe sample (blue contours) for IG; orange 
contours include also a Gaussian prior on the absolute magnitude $p\left(M_B\right)$.
The dotted and dashed contours include the correction included in Eq.~\eqref{eqn:mu_Ch} for 
P18 + BAO + SNe and P18 + BAO + SNe + $p\left(M_B\right)$, respectively.}
\label{fig:SNe_ig_Ch}
\end{figure*}

In Figs.~\ref{fig:SNe_ig_Ch_nobao} and \ref{fig:SNe_ig_Ch}, we compare the marginalized constraints on cosmological 
parameters for P18 + SNe  + $p(M_B)$ and P18 + BAO and P18 + BAO + SNe + $p(M_B)$ with and without the correction 
to the distance modulus relation introduced in Eq.~\eqref{eqn:mu_Ch}. 
We do not see any effect on means and uncertainties on any cosmological parameter for this combination of 
datasets. Note that for IG with $G_{\rm eff}(z=0) = G$, the difference on the redshift range of the Pantheon SNe 
sample ($0.01 < z < 2.3$) is smaller than 1\%.

\begin{figure*}
\centering
\includegraphics[width=0.98\textwidth]{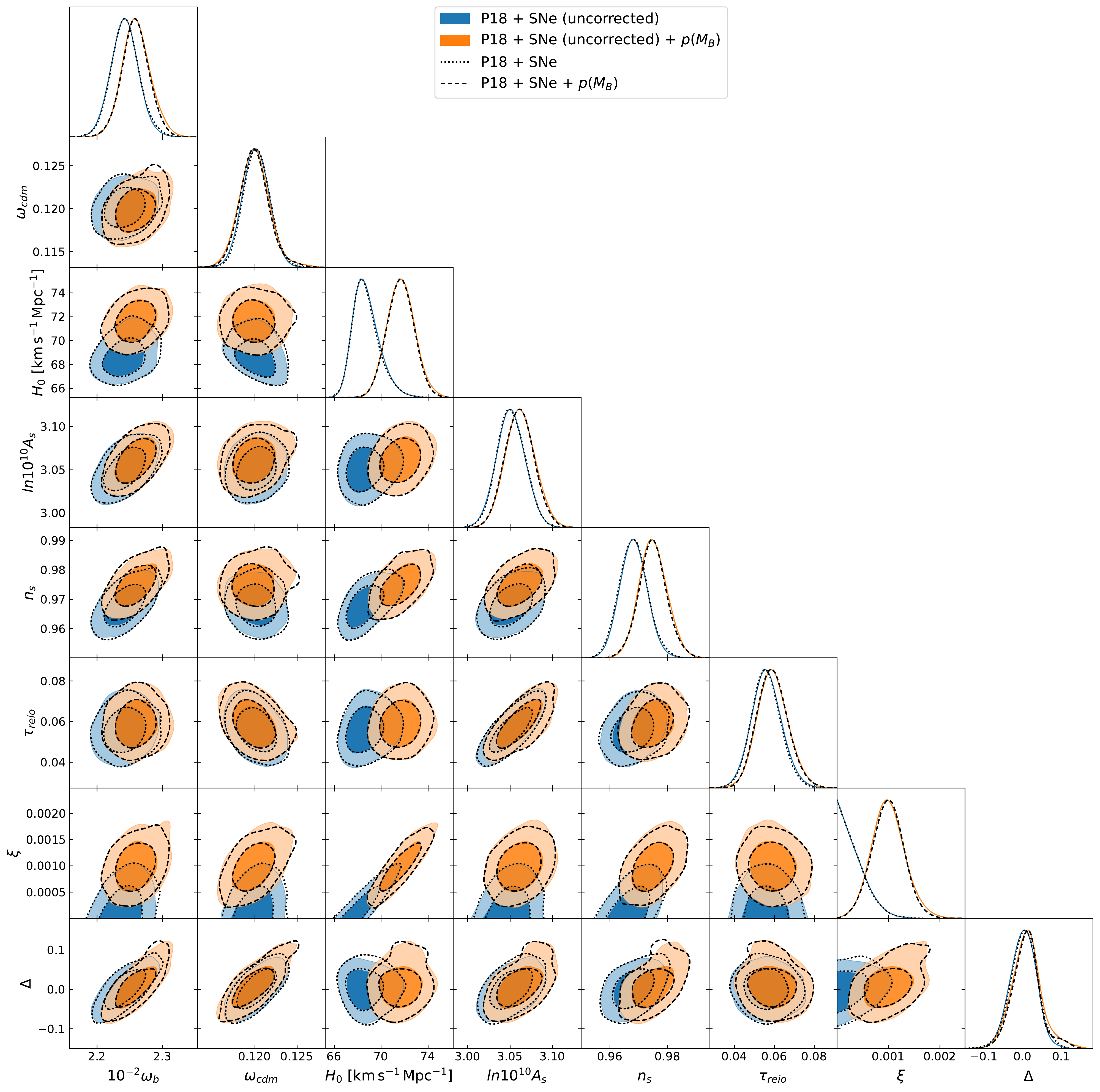}
\caption{As in Fig.\ref{fig:SNe_ig_Ch_nobao}, for IG with $\Delta$ allowed to vary.}
\label{fig:SNe_dig_Ch_nobao}
\end{figure*}

\begin{figure*}
\centering
\includegraphics[width=0.98\textwidth]{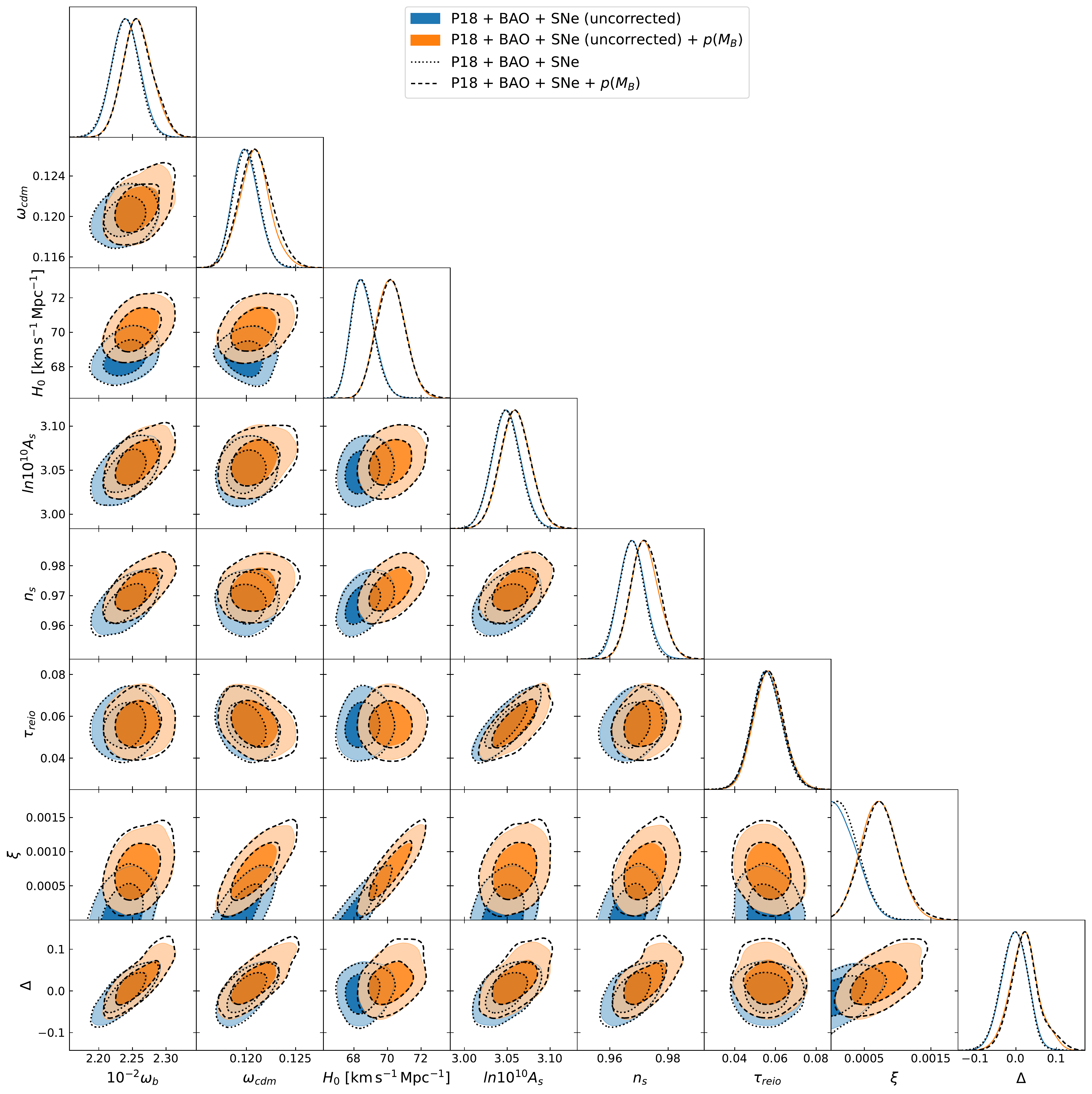}
\caption{As in Fig.\ref{fig:SNe_ig_Ch}, for IG with $\Delta$ allowed to vary.}
\label{fig:SNe_dig_Ch}
\end{figure*}

We find the same outcome also when we vary the imbalance $\Delta$, see 
Figs.~\ref{fig:SNe_dig_Ch_nobao} and \ref{fig:SNe_dig_Ch}. In this case the value of the Newton constant is 
constrained at 68\% CL to $G_{\rm eff}(z=0)/G = 1.05^{+0.05}_{-0.08}$ for P18 + BAO + SNe and to 
$G_{\rm eff}(z=0)/G = 1.04^{+0.06}_{-0.08}$ for P18 + BAO + SNe + $p(M_B)$ allowing for small correction from 
the distance modulus relation \eqref{eqn:mu_Ch}. Without including BAO data, we find at 68\% 
$G_{\rm eff}(z=0)/G = 1.005\pm 0.071$ for P18 + SNe and to 
$G_{\rm eff}(z=0)/G = 1.004\pm 0.070$ for P18 + SNe + $p(M_B)$.

\newpage
\footnotesize
\bibliographystyle{abbrv}


\end{document}